  \documentclass[aps,prd,groupedaddress,showpacs,notitlepage]{revtex4-1}
\usepackage{amsmath,amstext,amsbsy,amssymb}
\usepackage{bm}
\usepackage{color}

\newcommand{\Qre}{} %\Qre
\newcommand{\Qres}{} %\Qre^2

\newcommand{\ssB}{{\scriptscriptstyle{ B}}}

\newcommand{\ssM}{{\scriptscriptstyle{ M}}}

\newcommand{\sschi}{{\scriptscriptstyle{ \chi}}}

\newcommand{\ssmn}{{\scriptscriptstyle{\mu \nu }}}

\newcommand{\ssbfn}{{\scriptscriptstyle{(n)}}}
\newcommand{\ssbfu}{{\scriptscriptstyle{(u)}}}

\newcommand{\ssbfv}{{\scriptscriptstyle{(v)}}}
\newcommand{\ssbfI}{{\scriptscriptstyle{(I)}}}
\newcommand{\ssbfO}{{\scriptscriptstyle{(O)}}}
\newcommand{\ssbfM}{{\scriptscriptstyle{(M)}}}
\newcommand{\ssbfC}{{\scriptscriptstyle{(C)}}}

\long\def\symbolfootnote[#1]#2{\begingroup%
\def\thefootnote{\fnsymbol{footnote}}\footnote[#1]{#2}\endgroup}

\begin{document}
	
\title{Some Features  of  Semiclassical Chiral Transport  in Rotating Frames}

\author{\"{O}mer F. Dayi}
%\email{dayi@itu.edu.tr }
\author{Eda Kilin\c{c}arslan}
%\email{kilincarslan@itu.edu.tr }
\affiliation{%
	Physics Engineering Department, Faculty of Science and
	Letters, Istanbul Technical University,
	TR-34469, Maslak--Istanbul, Turkey}

\begin{abstract}
Semiclassical chiral kinetic theories in the presence of electromagnetic fields as well as vorticity can be constructed by means of some different  relativistic or nonrelativistic approaches. To cover the noninertial features of rotating frames one can start from the modified quantum kinetic equation of Wigner function in Minkowski spacetime. It provides a relativistic chiral transport equation whose nonrelativistic limit yields  a consistent  three-dimensional kinetic theory which does not depend explicitly on spatial coordinates. Recently a chiral transport equation in curved spacetime has been proposed and its nonrelativistic limit in rotating coordinates was considered   in the absence of electromagnetic fields. We  show that the modified theory can be extended to curved spacetime. The related particle current density and chiral transport equation for an inertial observer in the rotating frame are  derived. A novel three-dimensional chiral kinetic transport equation is established by  inspecting the nonrelativistic limit of the curved spacetime approach in the rotating frame for a comoving observer in the presence of electromagnetic fields. It explicitly depends on spatial coordinates. We prove that it is  consistent with the chiral anomaly, chiral magnetic and vortical effects. 

\end{abstract}

\maketitle

\section{Introduction}

In the presence  of external electromagnetic fields the charged, massless Dirac particles which can be  right- or  left-handed, exhibit  unusual features like the chiral magnetic effect \cite{kmw,fkw,kz} and  the chiral separation effect \cite{mz,jkr}. These manifest themselves in heavy-ion collisions \cite{kmw,fkw}. There are also some  similar phenomena due to rotation of  coordinate frame which affect dynamical behavior of chiral particles: The chiral vortical effect \cite{av,ss,ssz,bbbdls, lmb} and the local polarization effect \cite{lw,bpr,glpww}.

Dynamical features of chiral particles  have been investigated  mostly within the
semiclassical kinetic theories which offer  an intuitive understanding of their collective phenomena. Kinetic theories  can be studied either  in a manifestly covariant way in  Minkowski spacetime or within the nonrelativistic approach  where the physical content is apparent.   
Three-dimensional (3D) semiclassical chiral  transport theories  based on \cite{syD,sy}, are usually inspected  by  making  allowance for only the external electromagnetic fields or the rotation of coordinate frame. The first 3D  semiclassical kinetic   theory of chiral particles 
which takes into account the external electromagnetic fields as well as the rotation of the coordinate frame and 
consistent with the chiral anomaly 
was constructed in \cite{dky}. 
It generates the  chiral anomalous effects correctly as it should be.

A relativistic chiral kinetic theory (CKT) can be defined  \cite{hpy1, hpy2,hsjlz} starting from the quantum kinetic equation (QKE) obeyed by the Wigner function \cite{qBe,vge}.  However, it does not take into account all noninertial effects like the Coriolis force.  To overcome this  shortcoming QKE  was modified  by means of some frame dependent terms in \cite{dkR} . To have  better  insights into  the transport properties of particles, it is convenient to consider the 3D CKT generated by the relativistic chiral transport equation (CTE).  However, this limit is challenging, the nonrelativistic CKT should be consistent with chiral anomaly and anomalous currents.  The modified approach \cite{dkR}  gives rise to  a consistent 3D CKT which does not explicitly depend on the spatial coordinates   $\bm x,$  in the presence of both electrodynamic fields and vorticity. It is preferable to work with  a CTE which does not explicitly depend on the local positions of particles but global features like the angular velocity  $\bm \omega $ of the rotating frame. Nonetheless, explicit rotation velocity dependence is pertinent to  3D transport equations. In fact, the 3D CTE  established in \cite{dky}   depends explicitly on $\bm u =\bm x \times \bm \omega.$ 
This dependence is crucial in obtaining the continuity relation: If the rotation velocity dependent terms are suspended,  there will appear some undesired terms in the Liouville equation.

Relativistic chiral kinetic theories which we deal with are the ones provided by the  quantum kinetic equation without referring explicitly to the equilibrium distribution functions. In the approach of  \cite{cpww,gpw}  the  solution of quantum kinetic equation derived in \cite{glpww} was employed to define a CKT.  Yet another methods of furnishing chiral  transport equations were presented in  \cite{mv,cmt}.

 Recently CKT was studied in curved spacetime where the 3D limit in rotating coordinates has been exposed only for vanishing electromagnetic fields \cite{lgmh}.  Considering  QKE in curved spacetime \cite{fon} effectively means to modify it in a frame dependent manner. In fact the Coriolis force has been acquired by  choosing the metric adequate to deal with  rotating coordinates \cite{lgmh}. 
 The nonrelativistic limit in the absence of electromagnetic fields was discussed for two different choices of  four-velocities corresponding to  inertial and rotating observers.  For the latter choice the CTE which they obtained is the same with the one derived in \cite{dky} at $\bm x\approx 0.$  We would like to  understand how the formulations established in \cite{dky} and  \cite{dkR} are related to   the curved space formulation of chiral particles in rotating coordinates. We will first show that the modified QKE approach can be extended to curved spacetime.  We study it in rotating coordinates by choosing adequately  the metric tensor. The modification is relevant only  for an inertial observer as far as the centrifugal force is ignored.  We calculate the particle current density for an inertial observer in terms of the equilibrium distribution function resulting from two different choices of fluid four-velocities. Then  the CKE in the absence of electromagnetic fields is obtained. The phase space measure and first time derivatives of phase space variables which we acquire,  differ from the ones presented in \cite{lgmh}. In fact, CKT which  we acquire is partially in accord with the one established in  \cite{dky}.
 We then study the 3D limit of four-dimensional CTE for a rotating observer in the presence of  electromagnetic fields. The  CKT which follows possesses   $\bm x$ dependent terms and  resembles the one established in \cite{dky}. We show that it is consistent  with the chiral anomaly, the chiral vortical and magnetic effects. Hence it constitutes a new 3D CTE.  

In Sec. II we briefly review  the QKE and the existing modifications which take into account noninertial properties of rotating coordinates.  The modified kinetic equation in curved space is presented in   Sec. III. The particle current density and the resulting 3D CKT are obtained for an inertial observer.  Sec. IV is devoted to CKT for a rotating observer in the presence of external electromagnetic fields. We obtain a novel  3D CKT which  gives rise to the continuity equation and anomalous chiral effects correctly. In the last section the  results acquired and  future directions are discussed. 

\section{Wigner Function Formalisms}

Chiral theories which we  consider are  based on the quantum kinetic
equation 
\begin{equation}
\gamma^\mu \left(p_\mu + \frac{i\hbar}{2} \mathfrak{D}^{\ssbfI }_\mu\right) W(x,p) = 0.
\label{qkeO}
\end{equation}
$(x_\mu,p_\mu)$ are  the eight-dimensional phase space variables.  The superscript  $I=(O,M,C),$ standing for original, modified and curved,
indicates the different choices of derivative terms. 
 As it  will explicitly  be given shortly, 
 their differences lie in how they treat  the inertial properties of coordinate frame.
 $W(x,p)$  is the 
Wigner function for spin-$1/2$ fermions which can be  decomposed in terms of  the Clifford algebra  generators whose coefficients are
the scalar, pseudoscalar, vector,  axial-vector  and tensor fields.
We are interested in  the chiral vector fields given by the vector and axial-vector field  components  ${\cal V}_\mu $  and  ${\cal A}_\mu$ as
\begin{equation}
{\cal J}^\mu_\sschi = \frac{1}{2} ({\cal V}^\mu + \chi {\cal A}^\mu). \nonumber
\end{equation} 
$\chi =\pm 1,$ indicates the right- and left-handed  fermions. We take into consideration only the right-handed chiral vector field to expose our results, $ {\cal J}^\mu \equiv {\cal J}_1^\mu .$ 
The equations which they  obey  decouple from  the other components, leading to
\begin{eqnarray}
p_\mu {\cal J}^\mu & = & 0,
\label{1st,0} \\
\hbar \epsilon^{\mu \nu \alpha \rho} \mathfrak{D}^{\ssbfI }_\alpha {\cal J}_\rho&=& - 2 (p^\mu {\cal J}^{ \nu} -  p^\nu {\cal J}^{\mu}) ,
\label{2nd,0}\\
\mathfrak{D}^{\ssbfI }_\mu{\cal J}^{ \mu }& = & 0.
\label{3rd,0}
\end{eqnarray}

The original QKE  \cite{qBe,vge} was derived  starting from the Dirac equation  in Minkowski spacetime  yielding the derivative terms 
\begin{equation}
\mathfrak{D}^{\ssbfO}_\mu\equiv \nabla_\mu=\partial_\mu - \Qre F_{\mu\nu}   \partial^\nu_p . \label{nO}
\end{equation}
The derivatives with respect to the phase space variables denoted  $ \partial_\mu  \equiv \partial / \partial x^\mu,\   \partial^\mu_p  \equiv \partial / \partial p_\mu .$ We set $c=k=1$ as well as  
$Q =1 $ which is the electric charge of chiral particle
coupled to the external electromagnetic fields  described by $F_{\mu \nu}.$ 
For the sake of simplicity we deal with the electromagnetic field strength satisfying $\partial_\rho F_{\mu \nu}=0.$   

 In spite of the fact that the original QKE is covariant, it does not explicitly depend on the inertial properties of reference frame. The fluid vorticity or equivalently the angular velocity of the frame appears in its solution \cite{glpww}.  Therefore,
 the CTE designated by (\ref{nO}) does not lead to the Coriolis force. To surmount this disadvantage in  \cite{dkR}  the derivative terms of the original QKE is modified by means of the four-velocity $n^\mu$ of the frame satisfying $n_\mu n^\mu=1,$ as
\begin{equation}
\mathfrak{D}^{\ssbfM}_\mu\equiv \tilde{\nabla}_\mu^\ssbfn= \nabla_\mu
%\partial_\mu -\left[\Qre F_{\mu\nu} 
 +\left[ \partial_\nu n^\alpha   p_\alpha  n_\mu-\partial_\mu n^\alpha   p_\alpha  n_\nu  \right] \partial^\nu_p . \label{mm}
\end{equation}
We introduced the modification guided by the circulation tensor which provides the noninertial properties of fluids. 
This approach revealed to be essential in obtaining a  formulation  of 3D CTE which is not explicitly dependent of  the spatial coordinates $\bm x.$ In fact, it is the unique $\bm x$ independent 3D CKT consistent with the chiral anomaly  when both the external electromagnetic fields and fluid vorticity are present. Hence, as far as these properties are considered the modification terms seem to be unique.  This formalism possesses some similarities with the effective field theory approach \cite{ssz} but it is not possible to introduce a gauge field which generates the proposed  modification of QKE.

On the other hand frame dependent terms can also be incorporated in the original QKE by extending it to curved spacetime. This has been achieved by means of ``horizontal lift of the derivative operator in the cotangent bundle" \cite{fon,lgmh} yielding 
\begin{equation}
\mathfrak{D}^{\ssbfC}_\mu\equiv \Delta_{\mu}= \partial_\mu +\left[\Gamma^\lambda_{\mu\nu}p_\lambda -  \Qre F_{\mu\nu}  \right] \partial^\nu_p . \label{cot}
\end{equation}
$\Gamma^\lambda_{\mu\nu}$ denote the Christoffel symbols.  We suppressed spin connection because  it does not show up  in (\ref{1st,0})-(\ref{3rd,0}).

We are concerned with the semiclassical approximation where the Wigner function is expanded in Planck constant and the terms up to the first order are retained. In course of obtaining CKT one first solves (\ref{1st,0}) and (\ref{2nd,0}).   For $\nabla_\mu,$  they were solved in \cite{hpy1,hpy2,hsjlz}. As it has been demonstrated in \cite{lgmh}, this solution  can be extended to the curved spacetime by substituting $\nabla_\mu$ with $\Delta_\mu,$  giving
\begin{equation}
{\cal J}^{\mu} = p^\mu f \delta(p^2) + \hbar \Qre \tilde{F}^{\mu \nu}    p_\nu f \delta^\prime (p^2)   + \hbar S_\ssbfn^{\mu \nu} ({\Delta}_{\nu} f) \delta(p^2) ,
\label{generalform0}
\end{equation}
where
$ \delta^\prime (p^2) = d\delta(p^2)/dp^2.$ One  also expands  the  distribution function in $\hbar$ and  retain the first order: $f \equiv  f^{0} +\hbar  f^{1} . $   The spin tensor 
$$
S^{\mu \nu}_\ssbfn=
\frac{\epsilon^{\mu \nu \rho \sigma} p_\rho  n_\sigma  }{ 2 n \cdot p}  ,
$$ 
and 
the dual electromagnetic  field strength $\tilde{F}^{\mu \nu}  =\epsilon^{\mu \nu \alpha \rho}F_{\alpha \rho}/2,$ are introduced. By making use of (\ref{generalform0}) in the remaining equation (\ref{3rd,0}) one attains the relativistic CKT in curved spacetime:
\begin{equation}
\delta \Big( p^2 + \frac{  \hbar n_\alpha \tilde{F}^{\alpha \beta}   p_\beta}{n \cdot p} \Big)   
\Bigg[ p \cdot \Delta
+ \frac{ \hbar \Qre n_\mu \tilde{F}^{\mu \nu} \Delta_\nu}{n \cdot p}   +\hbar \Delta_\mu S_\ssbfn^{\mu \nu}\Delta_\nu \Big]f=0. \label{4deq0}
\end{equation}

We would like to examine the behavior of  massless Dirac particles in rotating coordinates. Metric components of 
the coordinate frame which  rotates  with the constant angular velocity $\bm \omega$ 
were given in   \cite{mtw,hn} as
\begin{eqnarray}
g_{00}=1-\bm u^2,\ \ g_{0i}=u^i,\ \ g_{ij}=-\delta_{ij}.\label{metric}
% \\ g^{00}=1,\ \ g^{0i}=g^{i0}=u^i,\ \ g^{ij}=-\delta_{ij}-u^iu^j. \nonumber
\end{eqnarray} 
We deal with nonrelativistic rotations, thus we let the rotation velocity   $u^i= \epsilon^{ijk}x^j\omega^k$  be  much smaller  than the speed of light:  $|\bm u| \ll  1.$  $\epsilon^{ijk}$ is totally antisymmetric and $\epsilon^{123}=1.$ 
 
The related Christoffel symbols can be seen to be
$$
\Gamma^{j}_{0i}=\Gamma^{j}_{i0}=\epsilon^{ijk} \omega ^k, \ \ \Gamma^i_{00}= \epsilon^{ijk} u^j \omega ^k,\ \ \Gamma^0_{\mu\nu}= \Gamma^i_{jk} =0.
$$
It yields  a vanishing   Riemann tensor. 

Following \cite{lgmh} we consider two different observers: The inertial observer with the four-velocity 
$$
u_{\mu}= (1,0,0,0),\ \ u^\mu =(1,u^i),
$$
and the rotating observer with the four-velocity
$$
v^\mu =\frac{1}{\sqrt{g_{00}}} \delta^\mu_0,\ \ v_{ \mu} =(\sqrt{g_{00}}, \frac{1}{\sqrt{g_{00} }} u^i ).
$$
The 3D CKT established in \cite{dky}  embraces  the Coriolis force as well as the centrifugal force. However, when one studies the 3D theories generated by the 4D CKTs the  centrifugal terms are ignored. Also here  we do not take into account  those effects which means that   $O({\bm u}^2)$  terms  and their derivatives with respect to spatial coordinates assumed to be vanishing. In accord with this assumption  we adopt the approximation
\begin{equation}
O(u^iu^j) \approx  0,\  O(u^i\omega^j) \approx  0. \label{nocf}
\end{equation}
It is worth mentioning  that this approximation is consistent with the fact that  $|\bm u| \ll  1.$ 

\section{Modified Kinetic Equation in Curved spacetime}
\label{mecs}

The modified  theory can be extended to the curved spacetime by substituting  $\mathfrak{D}^{\ssbfI}_\mu$ in (\ref{qkeO})  with 
\begin{equation}
{\Delta}_{\ssbfn \mu}= \partial_\mu +\left[\Gamma^\lambda_{\mu\nu}p_\lambda -  \Qre F_{\mu\nu}  + \partial_\mu n^\alpha   p_\alpha  n_\nu - \partial_\nu n^\alpha   p_\alpha  n_\mu \right] \partial^\nu_p .
\label{dmc}
\end{equation}
The signs of modification terms are altered with respect to (\ref{mm})  as it is convenient for  considering
 $p_\mu$ as  the  momentum variables.
The semiclassical 
solution of  (\ref{1st,0}) and (\ref{2nd,0})  can be attained as
\begin{eqnarray}
{\cal J}_\ssbfn^{\mu} &=&  \left(1-\frac{\hbar}{n \cdot p} S_\ssbfn^{\sigma \rho} g_{\rho \nu}( {\partial}_{ \sigma}n^\nu) \right) p^\mu f \delta(p^2) \nonumber\\
&&+ \hbar \Qre \tilde{F}^{\mu \nu}   p_\nu f \delta^\prime (p^2)  -  \hbar \epsilon^{\mu \nu \alpha \rho}  p_\nu (\partial_\alpha n^{\beta} ) p_\beta  n_\rho  f \delta^\prime (p^2) \nonumber\\
&&+ \hbar S^{\mu \nu}_\ssbfn ({\Delta}_{\ssbfn \nu} f^{0}) \delta(p^2).
\label{generalform}
\end{eqnarray}
We deal with the semiclassical approximation, so that  in (\ref{generalform}) the first order part of the distribution function $f^1,$ is arbitrary. In fact, we used this freedom  to write the second term in accord with the Minkowski spacetime formulation  \cite{dkR}. It is worth noting that to determine the distribution function completely one should solve  the remaining equation  (\ref{3rd,0}). 

We first would like to discuss 3D currents resulting from (\ref{generalform}) in rotating coordinates by choosing $n_\mu$ appropriately. Because of ignoring the centrifugal effects we deal with  $\bm u$  satisfying (\ref{nocf}). Then,   the unique possibility is to choose $n_\mu=u_\mu,$ because for $n_\mu=v_\mu$  the modification terms in (\ref{dmc}) vanish.

The zeroth component of (\ref{generalform}) yields particle number density and the spatial components  are used to define the chiral particle current density as
$$
j^i_{\ssbfu} =\int \frac{d^4p}{4\pi^3 \hbar^{3}} {\cal J}^{i}_{\ssbfu} .
$$ 
We would like to emphasize the fact that  terms which are at most linear in $\hbar$ are kept, obviously up to the    $\hbar^{-3}$ factor which is due to the definition of momentum space volume. 
Its partial integration reads 
\begin{eqnarray}
j_{\ssbfu}^{\mu} &=& \int \frac{d^4p}{4\pi^3 \hbar^{3}} \Big[ p^\mu f  - \hbar \Qre \partial_p^0\left(\tilde{F}^{\mu \nu}   p_\nu f/2p_0  \right) \nonumber\\
&&+\hbar \epsilon^{\mu \nu \alpha \rho} \partial_p^0\left(p_\nu (\partial_\alpha u^{\beta} ) p_\beta  u_\rho  f/2p_0 \right) \nonumber\\
&&
-\frac{\hbar}{u \cdot p} p^\mu S_\ssbfu^{\lambda \rho} g_{\rho \nu}\left( {\partial}_{ \lambda}u^\nu\right) f^0
+ \hbar S^{\mu \nu}_\ssbfu({\Delta}_{\ssbfu \nu} f^{0}) \Big] \delta(p^2). \nonumber
\end{eqnarray}

We adopt the equilibrium distribution function for rotating coordinates  as it  is given in \cite{css, lgmh}:
\begin{equation}
f_{eq}(x,p)= \left[1 +e^{ \left(p\cdot U-\mu +\frac{\hbar}{2} S^{\mu \nu}_\ssbfn\partial_\mu U_\nu\right)/T}\right]^{-1}. \label{feq}
\end{equation}
$U_\mu$ is the  four-velocity of  fluid and $n_\mu=u_\mu.$ 

One can first perform the  $p_0$ integral by solving the mass-shell condition $p^2=0$ as
$$
p_0=\pm |\bm p|-p_iu^i  ,
$$
where $|\bm p|=\sqrt{p_ip_i}.$ % We deal only with the particle solution. 
Then we write
$$
\delta (p^2) =\frac{\theta (p_0) \delta (p_0- |\bm p|+p_iu^i)}{ 2|\bm p|} +\frac{\theta (-p_0) \delta (p_0+ |\bm p|+p_iu^i)}{ 2|\bm p|} .
$$
We only display the positive part explicitly.  After integrating over $p_0,$ in the vicinity of $\bm x \approx 0$ the  current density can be written as
\begin{eqnarray}
j^i_\ssbfu &= &\int \frac{d^3p}{(2\pi\hbar)^3}\Big[-p_i +\frac{\hbar }{2 |\bm p|} \omega^i -\frac{3\hbar }{2|\bm p|^3}\omega^j p_j p_i\nonumber \\
& &
+\frac{\hbar \Qre}{2|\bm p|}\epsilon^{ijk}F_{jk}\frac{\partial}{\partial |\bm p|}+ \frac{\hbar \Qre}{2|\bm p|^3}\epsilon^{ijk}p_jF_{0k}\nonumber \\
& &
+ \frac{\hbar}{2 |\bm p|^2}\epsilon^{ijk} p_j \frac{\partial}{\partial x_k}\Big] f_{eq}|_{p_0=|\bm p|-p_iu^i } .\label{jmo}
\end{eqnarray}
Observe that $p_i/2 |\bm p|^3$  is the Berry curvature.

There are two possibilities of choosing the fluid four-velocity in the equilibrium distribution (\ref{feq}). For  
 $U_\mu=u_\mu,$   contributions arising from the second and third terms in (\ref{jmo}) cancel each other, so that  the chiral vortical effect is not generated.   (\ref{jmo}) only  leads to the magnetic current
$$
\bm j_\ssbfu^\ssB=\frac{\Qre\mu}{4\pi^2 \hbar^2} \bm B,
$$
where $\frac{1}{2}\epsilon^{ijk}F_{jk}=-B^i.$ It provides the chiral magnetic effect.
  The other possibility is to set
$U_\mu=v_\mu ,$ which  yields $\bm j_\ssbfu=\bm j_\ssbfu^\ssB + \bm j_\ssbfu^\omega,$ where
$$
\bm j_\ssbfu^\omega=\left(\frac{\mu^2}{2}+\frac{T^2\pi^2}{6}\right)\frac{\bm \omega }{2\pi^2\hbar^2},
$$
describes chiral vortical effect correctly. Obviously,
to  work out the momentum  integrals both  particle and antiparticle contributions should be taken into account. Although angular velocity dependence of the chiral current obtained in \cite{lgmh} differs from (\ref{jmo}),  the  results regarding chiral vortical effect are consistent.

Let us switch off the external electromagnetic fields and plug (\ref{generalform}) into the remaining Eq. (\ref{3rd,0}). For the sake of simplicity we set terms which are second order in angular velocity into zero and work in the vicinity of $\bm x \approx 0.$ After integrating over $p_0$ we obtain
\begin{equation}
\left(\sqrt{\kappa}\frac{\partial   }{\partial t} +\sqrt{\kappa}\dot{x}^i\frac{\partial }{\partial x^i}+ \sqrt{\kappa} \dot{p}_i \frac{\partial }{\partial p_i} \right)f(t,\bm x,\bm p)=0,\label{cttm} 
\end{equation}
where $f(t,\bm x,\bm p)\equiv f_{eq}(x,p_i,p_0=|\bm p|-p_iu^i).$ The phase space measure and the first time derivatives of phase space variables are  
\begin{eqnarray}
&\sqrt{\kappa}   =  1&+\frac{p_i\omega^i}{|\bm p|^2},  \label{ka1} \\
&\sqrt{\kappa} \dot{x}^i =  &- \mathrm{v}^i- \frac{\hbar \hat{ p}_j\omega^j \hat{ p}_i }{|\bm p|} , \label{ka2}\\
&\sqrt{\kappa} \dot{p}_i  =  & 2|\bm p|\epsilon^{ijk}\mathrm{v}^j\omega^k . \label{ka3}
\end{eqnarray}
We introduced the ``canonical velocity" 
 $$\mathrm{v}^i=\hat{ p}_i -\frac{\hbar \omega^i}{2|\bm p|}  +\frac{\hbar \hat{ p}_j\omega^j \hat{ p}_i }{2|\bm p|} .$$
(\ref{ka1}) and (\ref{ka3}) are in accord with the CKT derived in \cite{dky}. 
Let us compare (\ref{cttm})-(\ref{ka3}) with the CKT obtained  for inertial fluid in \cite{lgmh} where  phase space measure is 1 and $\ \dot{x}^i =-\hat{ p}_i  +u^i ,\ \dot{p}_i  =\epsilon^{ijk} p_j\omega^k .$  
First of all, some  versions of CKT are  related by phase space coordinate transformations \cite{sadofyev}, 
seemingly this is not the case for the  formalisms which we compare.
In the latter approach phase space measure and $\dot{{\bm x}}$ at $\bm x \approx 0, $  do not possess angular velocity dependence and in contrary to (\ref{ka3}) Coriolis force with the factor of 2 is generated if
one does not suppress the $\bm x$ dependence.  
 In fact, the latter formalism is  classical but the one derived here possesses  quantum corrections. This fact will be important especially when one considers collisions. Moreover, in Minkowski spacetime the modification of QKE is necessary to obtain a Coriolis like term and a satisfactory $\bm x$ dependent 3D CKT in the presence of electromagnetic fields. 

\section{3D CKT in  the Presence of Electromagnetic Fields}
\label{3D}

In this section we let   the particle 4-velocity be $U_\mu=v_\mu$ and consider the 4D CKT  in the comoving frame  $n_\mu=v_\mu ,$   provided by (\ref{4deq0}) as
\begin{equation}
\delta ( p^2 + \frac{\hbar \Qre  v_\alpha \tilde{F}^{\alpha \beta}   p_\beta}{v \cdot p})   
\{ p \cdot \Delta
+  \frac{ \hbar\Qre v_\mu \tilde{F}^{\mu \nu} \Delta_\nu }{v \cdot p}   +\hbar\Delta_\mu S^{\mu \nu}_\ssbfv  \Delta_\nu \}f=0. \label{4deq}
\end{equation}
In \cite{lgmh} this transport equation was studied in the absence of  electromagnetic fields. They showed that working with $p^\mu=g^{\mu \nu}p_\nu ,$ suits well in obtaining the 3D CKT which preserves the symmetry between magnetic field and angular velocity.  For this choice of momentum variables (\ref{cot})  is expressed as
$$
\Delta_{\mu}= \partial_\mu -\left[\Gamma^\nu_{\mu\lambda}p^\lambda +  \Qre F_{\mu\lambda} g^{\lambda \nu} \right] \partial_{p\,\nu}.
$$
We would like to study (\ref{4deq}) in the presence of the external electric and magnetic fields defined by
 $F_{0i}=(\bm E -\bm u\times \bm B)^i$ and $\frac{1}{2}\epsilon^{ijk}F_{jk}=-B^i.$  Note that in a frame rotating with the angular velocity $\bm \omega ,$ Maxwell equations are given in terms of $\bm B$ and $\bm E^\prime=\bm E -\bm u\times \bm B.$

To derive the 3D CKT we would like to integrate (\ref{4deq}) over  $p^0.$ To this aim we need to solve
$$
p^2 + \hbar \Qre \frac{v_\mu \tilde{F}^{\ssmn}   p_\nu }{v \cdot p}=0,
$$
for $p^0$ in terms of $p^i.$  By adopting the notation of \cite{lgmh} let us introduce  $\bm q =(p^1,p^2,p^3).$ The mass-shell condition can be solved as 
\begin{equation}
p^0=\pm\frac{|\bm q|}{\sqrt{g_{00}}}\sqrt{1+\frac{(\bm q\cdot \bm u)^2}{\bm q^2}+\hbar \Qre\frac{F_{\mu \nu} S^{\mu \nu}_\ssbfv }{\bm q^2}} - \frac{\bm q\cdot \bm u}{g_{00}} \label{fp0} .
\end{equation}
 Centrifugal terms are not taken into account, hence the rotation velocity satisfies  (\ref{nocf}). Moreover, we deal with weak external fields, so that we set  $O(E^2) \approx 0,$ and let the angular velocity and the magnetic field be in the same direction: $\bm \omega  \times \bm B=0.$  Therefore, in the semiclassical approximation, (\ref{fp0}) yields
\begin{equation}
p^0=\epsilon_q^+=|\bm q| - \bm q\cdot \bm u  -\hbar \Qre  |\bm q|\bm b \cdot \bm B^\prime ,
\label{p0}
\end{equation}
where 
$
 \bm{b}=\bm q/2|\bm q|^3
$
is the  Berry curvature and $\bm B^\prime =\bm B-\bm E\times \bm u $ is the external magnetic field observed in the coordinate frame moving with  the velocity $\bm \omega \times \bm x,$ given  by the Lorentz transformation with the Lorentz factor  $\gamma=1/\sqrt{1-\bm u^2} \approx 1.$ 
Observe that this is the mass-shell condition which we should employ in performing $p^0$ integrals. As we will see the effective  dispersion relation arising in the 3D CKT is independent of the electric field.
We only exhibit the particle solution, although antiparticles are essential to perform momentum integrals. 
The positive part of the delta function becomes
$$
\delta^+ (p^2 + \hbar \Qre \frac{v_\mu \tilde{F}^{\ssmn}   p_\nu }{v \cdot p}) =\frac{ \delta (p^0- \epsilon_q^+)}{ 2|\bm q|} 
\left[1- 2\hbar \Qre  \bm b \cdot \left( \bm E\times \bm u -\bm B \right)\right] .
$$
Now, by  integrating  (\ref{4deq}) over $p^0,$  one  acquires 
\begin{equation}
\left( \sqrt{\eta} \frac{\partial }{\partial t} + \sqrt{\eta} \dot{{\bm x}} \cdot \frac{\partial }{\partial \bm{x}} + \sqrt{\eta}  \dot{\bm q} \cdot\frac{\partial }{\partial \bm{q}}+I_0 \frac{\partial}{\partial p^0}\right) f|_{p^0=\epsilon_q^+}=0. \label{beq0}
\end{equation}
The phase space measure and the first time derivatives of phase space variables are
\begin{eqnarray}
&\sqrt{\eta} =1+ &  \ \hbar \bm{b}\cdot (\Qre\bm{B} + 2|\bm q|\bm{\omega } )
- \bm \nu\cdot \bm u ,  \label{S1} \\
&\sqrt{\eta} \cdot \dot{{\bm x}}=&{\bm \nu} + \Qre\bm E^\prime\times \bm b 
+   \hat{\bm q}  \cdot \bm b(\Qre\bm B+ 2 |\bm q| \bm \omega ) \nonumber \\
&&
+ 2 \Qre \hat{\bm q}  \ \bm b \cdot  [ \bm u \times \bm E] 
,\label{smlx} \label{S2}\\
&\sqrt{\eta} \cdot \dot{{\bm q}}=& \Qre\bm E^\prime+ {\bm \nu} \times (\Qre\bm{B} + 2{\cal E} \bm{\omega }) + \hbar \Qre  \bm E \cdot (\Qre \bm{B} + |\bm q| \bm{\omega }) \bm b \nonumber \\
&&+\hbar \Qre|\bm q| \bm b \cdot \bm \omega \bm E
-\Qre [ \bm u \times \bm E] \times {\bm \nu}  \nonumber \\
&& +\hbar \Qres\hat{\bm q}\cdot \bm b \bm E\cdot \bm B\bm u
+ 2 \Qres  \bm E\cdot (\hat{\bm q} \times \bm u) \bm b \times \bm B 
.  \label{smlp} \label{S3} 
\end{eqnarray}
We introduced the ``canonical velocity"  ${\bm \nu}=\partial {\cal E}/\partial \bm q$ with 
\begin{equation}
%\label{disp}
{\cal E} =q  -  \hbar q^2 ( \bm{b}\cdot \bm{\omega }) - \hbar \Qre q ( \bm{b}\cdot \bm B). \nonumber
\end{equation}
Observe that it  is the semiclassical  dispersion relation of a right-handed Weyl particle subject to the external electromagnetic fields   in rotating coordinates \cite{dky}. 
When electromagnetic fields are present  the coefficient of  $\partial f/\partial p^0$  in (\ref{4deq}) does not vanish. Its integral over $p^0$ 
turns out to be
\begin{eqnarray}
&I_0=&\Qre\hat{\bm q}\cdot \bm E^\prime \left( 1+ 2 \hbar \Qre \bm b \cdot \bm B+\hbar | \bm q | \bm b \cdot \bm \omega \right) 
+\hbar \Qre \bm b\cdot \bm q \bm E \cdot \bm \omega  \nonumber\\
&&-\Qre\bm E \cdot \bm u  - (1+ 2 \Qre \hbar \bm b \cdot \bm B) \Qre\hat{\bm q} \cdot (\bm B \times \bm u)
-\hbar \Qres\bm b \cdot \bm u \bm E \cdot \bm B . \nonumber
\end{eqnarray}
 However,  one can show that  it can be expressed as 
 $$
 I_0=  \sqrt{\eta}\dot{\bm q}\cdot (\partial  {\epsilon_q^+} /\partial \bm q).
 $$
Hence,  when we integrate  (\ref{4deq})  over $p^0,$ the coefficients of   $\partial f/\partial p^\mu,$ namely the last two terms of (\ref{beq0}), lead to
 \begin{eqnarray}
 & &\sqrt{\eta}\dot{\bm q} \cdot \left[ \frac{\partial f(x, p) }{\partial \bm q}\right]_{p^0= {\epsilon_q^+}} + 
 \sqrt{\eta}\dot{\bm q}\cdot \frac{\partial  {\epsilon_q^+}}{\partial \bm q} \left[\frac{\partial f( x,q) }{\partial p^0}\right]_{p^0= {\epsilon_q^+}} \nonumber \\
& & =  \sqrt{\eta}\dot{\bm q}\  \frac{\partial f(t,\bm x,    {\epsilon_q^+}, \bm q) }{\partial \bm q}.  \nonumber
 \end{eqnarray}
Therefore, we establish  the 3D CTE as
\begin{equation}
\left( \sqrt{\eta} \frac{\partial }{\partial t} + \sqrt{\eta} \dot{{\bm x}} \cdot \frac{\partial }{\partial \bm{x}} + \sqrt{\eta}  \dot{\bm q} \cdot\frac{\partial }{\partial \bm{q}}\right) f (t,\bm x,\bm q)=0, \label{beq}
\end{equation}
where $f (t,\bm x,\bm q) \equiv f(t,\bm x, p^0=\epsilon_q^+, \bm q).$ 

By making use of (\ref{S1})-(\ref{S3})  and the Maxwell equations in rotating coordinates one can show that the Liouville equation satisfied by the measure is
\begin{equation}
  \frac{\partial }{\partial t}\sqrt{\eta} +   \frac{\partial }{\partial \bm{x}} \left(\sqrt{\eta} \dot{{\bm x}}\right)+ \frac{\partial }{\partial \bm{q}}\left(\sqrt{\eta}  \dot{\bm q}\right)= \left(2\pi  \delta(\bm q) +2 \hat{\bm b}\cdot \bm u \right) \Qres\bm E\cdot \bm B .\label{Leq}
\end{equation}

In 3D  the chiral particle number and current densities are defined as 
\begin{eqnarray}
n  & = &  \int [dq] \sqrt{\eta}  f,\label{nil} \\
\bm j & = & \int[dq]\sqrt{\eta} \cdot \dot{\bm x} f  + \bm j_{\ssM} , \label{jil}
\end{eqnarray}
where $[dq] =d^3q/(2\pi\hbar)^3$ and
\begin{equation}
\label{curl}
\bm j_{\ssM }=\bm \nabla \times  \int[dq]  \hbar {\cal E} \bm b f ,
\end{equation}
is the magnetization current \cite{syD,cssyy,cipy,hpy1}.

By making use of (\ref{beq}) and (\ref{Leq}), one can show that the 4-divergence of the  4-current $(n,\bm j)$ yields 
the continuity equation with source:
\begin{equation} 
\label{ceqD0}
\frac{\partial n}{\partial t} + \bm {\nabla} \cdot \bm j = \frac{ \Qres  \bm{{ E}}\cdot \bm{B}}{(2\pi\hbar)^2}   f|_{\bm q =0}   .
\end{equation}
Note that on the  right-hand side only $ f^0$ appears. This continuity equation is consistent with the chiral anomaly.

Let us now focus on the currents which are proportional to  $\bm B$ and $\bm \omega,$  by  setting $\bm E=0.$  For $p^0=\epsilon_q^+,$ the equilibrium distribution function (\ref{feq}) with $n_\mu=U_\mu=v_\mu ,$ becomes
$$
f(t,\bm x, \bm p)=\frac{1}{e^{({\cal E}-\mu )/T}+1}.
$$
By plugging it into (\ref{jil}) and employing (\ref{S2}) one attains the chiral magnetic and vortical effects correctly: 
$$
\bm j =  \frac{ \left(3\mu^2+T^2\pi^2\right)}{12\pi^2\hbar^2} \bm \omega +\frac{\Qre\mu}{4\pi^2 \hbar^2} \bm B.
$$
To calculate the integrals we have taken into account both particle and antiparticle contributions.  We conclude that  (\ref{S1})-(\ref{beq})
describe a consistent  3D CKT.

\section{Discussions}

 In Minkowski spacetime the modified chiral  QKE   generates a consistent  3D CKT which does not explicitly depend on spatial coordinates. In Sec. \ref{mecs}  we studied its curved spacetime formulation  by ignoring  centrifugal force terms imposing  the conditions (\ref{nocf}).  The  modification terms survive for the  inertial observer: $n_\mu=u_\mu.$  We derived  the current density at $\bm x \approx 0$ and show that the chiral vortical effect is   generated  for  the particle (fluid) velocity $U_\mu=v_\mu.$ For  $U_\mu=u_\mu$  the current density does not lead to any  angular momentum dependent term. This is not surprising because the equilibrium distribution function  of rotating fluid in Minkowski spacetime \cite{css} is consistent only with the former choice.  We derived  the transport equation in the absence of electromagnetic fields as in (\ref{cttm}). It furnishes the phase space measure and  velocities which in part coincide with the ones obtained in \cite{dky} at $\bm x \approx 0.$  Hence, we can conclude that the modified QKE in curved spacetime is needed to acquire the phase space measure and the first time derivatives of phase space variables correctly for the  inertial observer.
 
In Sec. \ref{3D}  the novel CTE (\ref{beq})  is established for $n_\mu =v_\mu .$ It is similar to the CKT obtained directly in 3D  \cite{dky}. The difference is mainly in the explicitly $\bm x$-dependent terms of (\ref{S1})-(\ref{S3}).  The 3D CKT of \cite{dky} was constructed starting from the scalar and  vector fields which can be associated with Coriolis and centrifugal forces experienced by a massive particle.  One can examine if a similar approach in 3D exists which can be associated with  the new 3D CKT.  Another open question is how to incorporate the centrifugal force in the 3D CKT starting with the 4D curved spacetime formulation of chiral particles.  It is a hard task, because one should not only keep terms at the order of $\bm u^2$ but also, at least initially, the $\bm{u}^3$ terms whose derivatives with respect to spatial coordinates are at the order of  $\bm u^2.$ This would also clarify if an underlying 3D construction mentioned above exists.

We only presented transport equations without collisions.
The explicitly $\bm x$-dependent CKT  of  \cite{dky} has been extended to cover collisions by adopting the relaxation time method \cite{ofdek}. This allowed us to  study nonlinear transport properties of chiral plasma.  The novel formalism constructed here can be studied in a similar manner. Collisions can be introduced  by means of  the relaxation time formalism and the particle current densities provided by them can be calculated. Once this is done one can compare  the particle  currents generated by these CKTs. This would serve as a testing ground  for  deciding which CKT suits better with the observable effects.

\begin{acknowledgments}
We would like to thank Xu-Guang Huang for the illuminative correspondence on  their work.
	This work is supported by the Scientific and Technological Research Council of Turkey (T\"{U}B\.{I}TAK) Grant No. 117F328.
	
\end{acknowledgments}
\newcommand{\PRL}{Phys. Rev. Lett. }
\newcommand{\PRB}{Phys. Rev. B }
\newcommand{\PRD}{Phys. Rev. D }

% \end{document}

\end{document}